\documentclass[pra,aps,twocolumn,superscriptaddress,showpacs,showkeys,nofootinbib]{revtex4-1}
\usepackage{amsmath,amsthm,amssymb,graphicx,subfigure,xcolor}
\usepackage{graphics,float}
\usepackage{color}
\usepackage[unicode=true]{hyperref}
\usepackage{booktabs}
\usepackage{tabularx}
\usepackage{tabu}
\usepackage{multirow}
\usepackage{graphicx}
\usepackage{subfigure}
\usepackage{setspace}
\usepackage{amsmath}
\makeatletter
\renewcommand*\env@matrix[1][\arraystretch]{%
  \edef\arraystretch{#1}%
  \hskip -\arraycolsep
  \let\@ifnextchar\new@ifnextchar
  \array{*\c@MaxMatrixCols c}}
\makeatother
\hypersetup{
     colorlinks=true,       		
     linkcolor=blue,          	
     citecolor=red,            
     urlcolor=magenta,           	
 }
\vspace{-0.8cm}
\newtheorem{theorem}{\indent Theorem}
\newtheorem*{theorem*}{\indent Theorem}

\newtheorem*{corollary*}{\indent Corollary}

\newtheorem*{lemma*}{\indent Lemma}
\newtheorem{proposition}{\indent Proposition}
\newtheorem*{proposition*}{\indent Proposition}
\theoremstyle{definition}
\newtheorem{definition}{\indent Definition}
\newtheorem*{definition*}{\indent Definition}
\theoremstyle{remark}

\newtheorem*{remark*}{\indent Remark}

\begin{document}

\title{Measures of imaginarity  and quantum state order}

\author{Qiang Chen}
\author{Ting Gao}
\email{gaoting@hebtu.edu.cn}
\affiliation{School of Mathematical Sciences, Hebei Normal University, Shijiazhuang 050024, China}
\author{Fengli Yan}
\email{flyan@hebtu.edu.cn}
\affiliation{College of Physics, Hebei Key Laboratory of Photophysics Research and Application, Hebei Normal University, Shijiazhuang 050024, China}

\begin{abstract}
Complex numbers are widely used in both classical and quantum physics, and  play an important role in describing quantum systems and their dynamical behavior.  In this paper we study several measures of imaginarity of quantum states in the framework of resource theory, such as the measures based on $l_{1}$ norm,  and convex function, etc. We also investigate  the influence of the quantum channels on quantum state order for a single-qubit.
\end{abstract}



\maketitle

\section{INTRODUCTION}

Quantum resource theory provides a  method for exploring the properties of quantum systems
 \cite{3,4}. In this  theory the resource of the quantum system is  quantified by  an operational method and the information processing tasks which  can be realized are determined by the resource consumed. For example, in the resource theory of entanglement, the quantization of entanglement  \cite{Wootters, 9,HongGaoYan,10, gao} and a series of applications of entanglement, such as  quantum  key distribution \cite{Ekert,Bennett3,33, 37, 50, 32, 39}, quantum teleportation \cite{Bennett1,11}, quantum direct communication \cite{LLT, YanZhang,13,14},   quantum secret sharing \cite{HBB,12} have been provided. In recent years, researchers proposed many resource theories, such as resource theories of coherence \cite{2,6,21}, asymmetry \cite{17}, quantum thermodynamics \cite{16}, nonlocality \cite{8}, superposition \cite{7}, etc. In addition, people  also have developed applicable quantities in mathematical framework of resource theory \cite{5}.

One feature of quantum mechanics is the use of imaginary numbers. Although imaginary numbers are used to describe the motion of an oscillatory  in classical physics, they play a very important role in quantum mechanics, because the wave functions of quantum system all involve complex numbers \cite{1}. Consider, for example, the polarization density matrix of a single photon
in the $\{|H\rangle, |V\rangle\}$ basis, where $|H\rangle$ and $|V\rangle$ express the horizontal polarization, and  vertical polarization,   respectively. As a matter of fact, the imaginary numbers in the density matrix cause the rotation of the electric field vector. Based on this phenomenon, Hickey and Gour \cite{18} came up with imaginarity resource theory. In this theory, the density matrix with imaginary  elements is defined as resource state, otherwise as free state. Hickey and Gour \cite{18} also defined  the largest class of free operations. For the special physical constraints, some free operations are obtained, and then the theoretical framework of imaginarity resource is established. In this framework, several measures of imaginarity  are given, and a state conversion  condition for the pure states of a single qubit is discussed. Furthermore, in 2021, Wu et al \cite{22,23}   proposed the robustness measure of  imaginarity, and gave the transformation condition of states  of a single qubit under free operation.

In this paper, we investigate several measures of imaginarity in the framework of resource theory.  The rest of this paper is organized as follows. In Sec. II, we review  some  concepts including the real states, the free operations and  measures of imaginarity. In Sec. III, we mainly study whether the  measures of imaginarity based on $l_{p}$ norm, $p-$norm, and  convex roof extended are good measures  in the framework of  the resource theory.  The   influence of the quantum channels on quantum state order for a single-qubit  is discussed in Sec. IV.

\section{Background}\label{sec:review}

\subsection{Theoretical framework of imaginarity resource }\label{sec:review}

 Suppose  $\{|j\rangle\}^{d-1}_{j=0}$ is a fixed basis in a $d$-dimensional Hilbert space $\mathcal{H}$. We use  $\mathcal{D}(\mathcal{H})$ to denote   the set of density operators acting on $\mathcal{H}$. In fact, a  quantum state  is  descibed by a density operator $\rho$ in $\mathcal{D}(\mathcal{H})$. The theoretical framework of imaginarity resource \cite{18}   consists of three ingredients: real states (free states),  free operations and  measures of imaginarity.  They   are defined as follows.

$\mathbf{Real~state}$ \cite{18,22,23}:  In a fixed basis $\{|j\rangle\}^{d-1}_{j=0}$,  if quantum state
\begin{equation}\label{7}
\begin{aligned}
\rho=\sum_{jk}\rho_{jk}|j\rangle\langle k|
\end{aligned}
\end{equation}
satisfies  each $\rho_{jk}\in\mathbb{R}$, we call $\rho$   a real state (free state). Here $\mathbb{R}$ is the set of real numbers. We denote the set of all real states by $\mathcal{F}$.

In other words, the density matrices of free states  are real with respect to a fixed basis.

$\mathbf{Free~operation}$ \cite{18}:  Let $\Lambda$ be a quantum operation with Kraus operators $\{K_j\}$, and $\rho$ be a density operator, $\Lambda[\rho]=\sum_{j}K_{j}\rho K_{j}^{\dagger}$. We say that $\Lambda$ is a free operation (real quantum operation) if
\begin{equation}\label{7}
\begin{aligned}
\langle i|K_{j}|l\rangle\in\mathbb{R}
\end{aligned}
\end{equation}
for arbitrary $j$ and $i,l\in \{0,1,\cdots, d-1\}$.

$\mathbf{Measures~of~imaginarity}$ \cite{22}: A measure of imaginarity is a function $M$ : $\mathcal{D}(\mathcal{H})\rightarrow[0,\infty)$ such that

1.~$M(\rho)=0$ if and only if $\rho\in\mathcal{F}$.

2.~$M(\varepsilon(\rho))\leq M(\rho)$,  where $\varepsilon$ is a free operation.

Condition 2 is also called monotonic.

It is easy to observe   that this theory is  basis dependent, the real states  do not possess any resource, and the free operations  can not generate resources from real states.

\section{Measures~of~imaginarity}

Let us begin to discuss the quantization of imaginarity, which plays a very important role in determining the resources of a given quantum state.

Two measures of imaginarity of the quantum state $\rho$ have been  proposed in \cite{18}. They are the measure of imaginarity based on the $1-$norm,
\begin{equation}\label{7}
\begin{aligned}
M(\rho)=\min_{\sigma\in\mathcal{F}}\|\rho-\sigma\|_{1}=\frac{1}{2}\|\rho-\rho^{\rm T}\|_{1},
\end{aligned}
\end{equation}
where $\rho^{\rm T}$ denotes the transposition of density matrix $\rho$, $\|A\|_{1}={\rm Tr}[(A^{\dagger}A)^{1/2}]$  is the  $1-$norm of matrix $A$ \cite{2},  and  the robustness of imaginarity
\begin{equation}\label{7}
\begin{aligned}
R(\rho)=\min_{\sigma\in\mathcal{D}(\mathcal{H})}\{s\geq0:\frac{s\sigma+\rho}{1+s}\in\mathcal{F}\}.
\end{aligned}
\end{equation}
The geometric measure of imaginarity for pure states $|\psi\rangle$ is \cite{ 22}
\begin{equation}\label{7}
\begin{aligned}
M_{\rm g}(|\psi\rangle)=1-\max_{|\phi\rangle\in\mathcal{F}}|\langle \phi| \psi\rangle|^{2}.
\end{aligned}
\end{equation}

Next we discuss several important distance-based imaginarity functions. Consider the function constructed based on the $l_{p}$ norm. The $l_{p}$ norm of a matrix $A$ \cite{2} is defined as
\begin{equation}\label{7}
\begin{aligned}
\|A\|_{l_{p}}=\{\sum_{ij}|A_{ij}|^{p}\}^{1/p}.
\end{aligned}
\end{equation}

Specially we can  define the function  based on the $l_{1}$ norm as
\begin{equation}
M_{l_{1}}(\rho)=\min_{\sigma\in\mathcal{F}}\|\rho-\sigma\|_{l_{1}},
\end{equation}
 where $\rho$ is an arbitrary quantum state, $\sigma\in\mathcal{F}$ is real quantum state.    Then one can obtain the following result.

\begin{theorem} $M_{l_{1}}(\rho)=\sum_{i\neq j}|{\rm Im}(\rho_{ij})|$, and  $M_{l_{1}}(\rho)$ is a measure of imaginarity for free operations being all real operations within complete positivity trace-preserving (CPTP) quantum operations, where ${\rm Im}(\rho_{ij})$ represents the imaginary part of the matrix element $\rho_{ij}$.
\end{theorem}

\textit{\textbf{Proof.}} Firstly, we  prove $M_{l_{1}}(\rho)=\sum_{i\neq j}|{\rm Im}(\rho_{ij})|$. Obviously, each quantum state $\rho=(\rho_{ij})$ in a $d$-dimensional Hilbert space can be written as $\rho=(\rho_{ij})=(a_{ij}+{\rm i}b_{ij})$, where  $a_{ij},b_{ij}$ are real numbers, and when $i=j$,   then $b_{ij}=0$ holds. The real state $\sigma=(\sigma_{ij})=(c_{ij})$ with   $c_{ij}$  being real numbers.  Hence
\begin{equation}\label{7}
\begin{aligned}
&\quad\|\rho-\sigma\|_{l_{1}}\\
&=|(a_{11}-c_{11})|+|(a_{22}-c_{22})|+\cdots +|(a_{dd}-c_{dd})|\\
&\quad+2|(a_{12}-c_{12})+b_{12}{\rm i}|+2|(a_{13}-c_{13})+b_{13}{\rm i}|\\
&\quad+\cdots+2|(a_{1d}-c_{1d})+b_{1d}{\rm i}|\\
&\quad+2|(a_{23}-c_{23})+b_{23}{\rm i}|+2|(a_{24}-c_{24})+b_{24}{\rm i}|\\
&\quad+\cdots+2|(a_{2d}-c_{2d})+b_{2d}{\rm i}|\\
&\quad+2|(a_{(d-1)d}-c_{(d-1)d})+b_{(d-1)d}{\rm i}|\\
&=\sum_{ij}\sqrt{(a_{ij}-c_{ij})^{2}+b_{ij}^{2}}.
\end{aligned}
\end{equation}
 Clearly,   the minimum  of $\|\rho-\sigma\|_{l_{1}}$ occurs at  $c_{ij}=a_{ij}$.  That is, when $\sigma={\rm Re} (\rho)$, one gets
\begin{equation}\label{7}
\begin{aligned}
 M_{l_{1}}(\rho)=\sum_{i\neq j}|{\rm Im}(\rho_{ij})|.
\end{aligned}
\end{equation}
Here ${\rm Re} (\rho)$ stands for  the real part of $\rho$.  It means that  ${\rm Re} (\rho)$ is the closest real state of $\rho$.

New we demonstrate that the function $M_{l_{1}}(\rho)$  is a measure of imaginarity of quantum state $\rho$.

Obviously, for an arbitrary  quantum state $\rho$, we have
\begin{equation}\label{7}
\begin{aligned}
M_{l_{1}}(\rho)=\sum_{i\neq j}|{\rm Im}(\rho_{ij})|\geq0.
\end{aligned}
\end{equation}
For a real quantum state $\rho$ we can easily derive $M_{l_{1}}(\rho)=0$ by Eq.(\ref{7}).

When the function $M_{l_{1}}(\rho)=0$, one  has
\begin{equation}\label{8}
\begin{aligned}
|{\rm Im}(\rho_{ij})|=0.
\end{aligned}
\end{equation}
It implies that the matrix elements of quantum state $\rho$ are  real numbers. Hence  quantum state $\rho$ is real.

After that, we want to show that $M_{l_{1}}(\rho)$ is  monotonic under an arbitrary  real operation within CPTP.

Assume  $\varepsilon$ is the real operation within CPTP, $\rho$ and $\sigma$ are two density operators, according to the definition of $l_{1}$ norm \cite{2}, we have
\begin{equation}\label{8}
\begin{aligned}
\|\varepsilon (\rho )-\varepsilon(\sigma)\|_{l_{1}}\leq\|\rho-\sigma\|_{l_{1}}.
\end{aligned}
\end{equation}

Evidently, a quantum state $\rho$ can be written as $\rho=\rho_{{\rm R}}+{\rm i}\rho_{{\rm I}}$, where $ \rho_{{\rm R}}=\frac{1}{2}( \rho+\rho^{{\rm T}} ),\rho_{{\rm I}}=\frac{1}{2{\rm i}} (\rho -\rho^{{\rm T}})$. It is not difficult to observe  that $\rho_{{\rm R}}$ is real symmetric, $\rho_{{\rm I}}$ is real antisymmetric, and
\begin{equation}\label{8}
\begin{aligned}
{\rm Tr}\rho_{{\rm R}}={\rm Tr}[\frac{1}{2}( \rho+\rho^{{\rm T}} )]=\frac{1}{2}[{\rm Tr}(\rho)+{\rm Tr}(\rho^{{\rm T}})]=1,
\end{aligned}
\end{equation}

\begin{equation}\label{8}
\begin{aligned}
\langle\psi|\rho_{{\rm R}}|\psi\rangle=\frac{1}{2}\langle\psi|\rho|\psi\rangle+\frac{1}{2}\langle\psi|\rho^{{\rm T}}|\psi\rangle\geq0.
\end{aligned}
\end{equation}
Therefore  $\rho_{\rm R}$ is the real density matrix. According to the $l_{1}$ norm of the matrix is contracted under CPTP, one can  obtain

\begin{equation}\label{8}
\begin{aligned}
&\quad M_{l_{1}}(\varepsilon(\rho))\\
&=\inf_{\sigma\in\mathcal{F}} \|\varepsilon(\rho)-\sigma\|_{l_{1}}\\
&\leq\|\varepsilon(\rho)-\varepsilon(\rho_{\rm R})\|_{l_{1}}\\
&=\|\varepsilon (\rho_{{\rm R}}+{\rm i}\rho_{{\rm I}} )-\varepsilon(\rho_{{\rm R}})\|_{l_{1}}\\
&\leq\|\rho-\rho_{{\rm R}}\|_{l_{1}}\\
&=\|{\rm i}\rho_{{\rm I}}\|_{l_{1}}\\
&=M_{l_{1}}(\rho).
\end{aligned}
\end{equation}
Thus  we arrive at   that the function $M_{l_{1}}(\rho)$  is a measure of imaginarity for free operations being all real operations within CPTP quantum operations. The proof of Theorem 1 has been completed.

However, for the functions   induced by the $l_{p}$  norm or $p-$norm \cite{6} we have the following conclusion.

\begin{theorem} For any quantum state $\rho$ in a $d$-dimensional Hilbert space, when $p>1$, both the function $M_{l_{p}}(\rho\otimes\frac{\mathbb{I}}{d})$ and function $M_{p}(\rho\otimes\frac{\mathbb{I}}{d})$ induced by the $l_{p}$  norm and  $p-$norm respectively,  do not satisfy monotonicity  under all real operations within CPTP mappings.
\end{theorem}

\textit{\textbf{Proof.}} It is not difficult to observe that  for a particular real state
\begin{equation}\label{9}
\begin{aligned}
\rho_{1}=|0\rangle\langle0|,
\end{aligned}
\end{equation}
there exists a real operation $\Lambda$ which transforms   the quantum state
\begin{equation}\label{8}
\begin{aligned}
\rho_{2}=\frac{\mathbb{I}}{d}
\end{aligned}
\end{equation}
to the quantum state $\rho_{1}$.  Here $\mathbb{I}$ is the $d$-dimensional identity operator, the Kraus operators of the real operation $\Lambda$ are $\{K_{i}=|0\rangle\langle i-1|\}$, and $\{K_{i}\}$ satisfy  $\sum_{i=1}^d K_{i}^{\dagger}K_{i}=\mathbb{I}$.

We choose the real operation $\tilde\Lambda$, whose Kraus operators are $\{\tilde{K_{i}}=\mathbb{I}\otimes K_{i}\}$.  Clearly,  $\{\tilde{K_{i}}\}$ satisfy  $\sum_{i}\tilde{K_{i}^{\dagger}}\tilde{K_{i}}=\mathbb{I'}$, where $\mathbb{I'}$ is the identity operator of the direct product space. Then we have

\begin{equation}\label{9}
\begin{aligned}
M_{l_{p}}(\tilde\Lambda[\rho\otimes\frac{\mathbb{I}}{d}]) &=M_{l_{p}}(\rho\otimes|0\rangle\langle0|)\\
&=\{\sum_{ij}|{\rm Im}(\rho\otimes|0\rangle\langle0|)_{ij}|^{p}\}^{1/p}\\
&=M_{l_{p}}(\rho)\\
&>M_{l_{p}}(\rho\otimes\frac{\mathbb{I}}{d}).
\end{aligned}
\end{equation}
Here ${\rm Im}(\rho\otimes|0\rangle\langle0|)_{ij}$ represents the imaginary part of the matrix element $(\rho\otimes|0\rangle\langle0|)_{ij}$ . The above inequality takes advantage of the following results
\begin{equation}\label{10}
\begin{aligned}
M_{_{l_{p}}}(\rho\otimes\frac{\mathbb{I}}{d})&=\{\sum_{ij}|{\rm Im}(\rho\otimes\frac{\mathbb{I}}{d})_{ij}|^{p}\}^{1/p}\\
&=d^{\frac{1}{p}-1}M_{_{l_{p}}}(\rho)\\
&<M_{_{l_{p}}}(\rho).\\
\end{aligned}
\end{equation}

Obviously, Eq.(\ref {9}) indicates  that   when $p>1$, function $M_{l_{p}}$ does not satisfy the condition $M_{l_{p}}(\varepsilon(\rho))\leq M_{l_{p}}(\rho)$  for arbitrary free operation $\varepsilon$ and quantum state $\rho$. That is  when $p>1$, function $M_{l_{p}}$ can  not be regarded as a measure of imaginarity.

For a  matrix $A$, its $p$-norm $\|A\|_p$   is defined as $[{\rm Tr}(A^+A)^{\frac{p}{2}}]^{\frac {1}{p}}$.  When $p>1$, for $p$-norm induced function
\begin{equation}\label{10}
\begin{aligned}
M_{p}(\rho)=\min_{\sigma\in\mathcal{F}}\|\rho-\sigma\|_{p},
\end{aligned}
\end{equation}
we have
\begin{equation}\label{10}
\begin{aligned}
M_{p}(\tilde\Lambda[\rho\otimes\frac{\mathbb{I}}{d}]) &=M_{p}(\rho\otimes|0\rangle\langle0|)\\
&=M_{p}(\rho)\\
&>M_{p}(\rho\otimes\frac{\mathbb{I}}{d}).
\end{aligned}
\end{equation}
The inequality above can be derived from
\begin{equation}\label{10}
\begin{aligned}
M_{p}(\rho\otimes\frac{\mathbb{I}}{d}) & \leq\min_{\sigma\in\mathcal{F}}\|\rho\otimes\frac{\mathbb{I}}{d}-\sigma\otimes\frac{\mathbb{I}}{d}\|_{p}\\
&=\min_{\sigma\in\mathcal{F}}\|(\rho-\sigma)\otimes\frac{\mathbb{I}}{d}\|_{p}\\
&=\min_{\sigma\in\mathcal{F}}\|\rho-\sigma\|_{p}\|\frac{\mathbb{I}}{d}\|_{p}\\
&=M_{p}(\rho)\|\frac{\mathbb{I}}{d}\|_{p}\\
&<M_{p}(\rho).
\end{aligned}
\end{equation}
Thus we have demonstrated that when $p>1$,  the function $M_{p}$ violates monotonicity  under all real operations within CPTP mappings.  Hence Theorem 2 is true.

Let us  discuss the measure of imaginarity based on relative entropy. In resource theory of coherence,  coherence measure $C_{{\rm r}}(\rho)$ based on relative entropy  satisfies the axiomatic condition of coherence measure \cite{4}, and its expression being similar to coherence distillation \cite{21} is
\begin{equation}\label{10}
\begin{aligned}
C_{{\rm r}}(\rho)=S(\Delta'(\rho))-S(\rho),
\end{aligned}
\end{equation}
where $\Delta'$ is the decoherence  operation and $S(\rho)$ stands for   Von Neumann entropy of quantum state $\rho$.

Similar to resource theory of coherence, here  we need  an  operator $\Delta$.
\begin{definition} The mathematical operator $\Delta$ is defined by
\begin{equation}\label{6}
\begin{aligned}
\Delta(\rho)=\frac{1}{2}(\rho+\rho^{\rm T}),
\end{aligned}
\end{equation}
where $\rho$ is any quantum state.
\end{definition}

Evidently, $\Delta$   is just a simple mathematical operator, rather than a free operation.
The relationship between  $\Delta(\rho)$ and quantum real operation satisfying the physically consistent condition \cite{18} can be stated as the following theorem.

 \begin{theorem} Let $\varepsilon$ be a  real operation within CPTP. If  $\varepsilon$  satisfies the  condition of physical consistency.  Then for any quantum state $\rho$, we have
 \begin{equation}\label{6}
\begin{aligned}
 \varepsilon(\Delta(\rho))=\Delta(\varepsilon(\rho)).
\end{aligned}
\end{equation}
\end{theorem}

\textit{\textbf{Proof.}} For any quantum state $\rho$, because the real operation $\varepsilon$ is linear, hence one has

 \begin{equation}\label{6}
\begin{aligned}
\varepsilon[\Delta(\rho)]&=\varepsilon[\frac{1}{2}(\rho+\rho^{{\rm T}})]\\
&=\frac{1}{2}[\varepsilon(\rho)+\varepsilon(\rho^{\rm T})]\\
&=\frac{1}{2}[\varepsilon(\rho)+\varepsilon(\rho)^{\rm T}]\\
&=\Delta(\varepsilon(\rho)),
\end{aligned}
\end{equation}
where the third equality of the above equation is obtained from the condition of physical consistency  \cite{18}. Therefore Theorem 3 holds.

The quantum relative entropy between quantum states $\rho$ and $\sigma$ is usually taken as \cite{20}
\begin{equation}\label{6}
\begin{aligned}
S(\rho\|\sigma)={\rm Tr}[\rho\log_{2}\rho]-{\rm Tr}[\rho\log_{2}\sigma].
\end{aligned}
\end{equation}
The relative entropy  of imaginarity of   a quantum state $\rho$  is defined as \cite{41}
\begin{equation}\label{6}
\begin{aligned}
M_{{\rm r}}(\rho)=\min_{\sigma\in\mathcal{F}} S(\rho\|\sigma).
\end{aligned}
\end{equation}
Then the relative entropy function $M_{{\rm r}}(\rho)$ can be reexpressed as \cite{41}

\begin{equation}\label{6}
\begin{aligned}
 M_{{\rm r}}(\rho)=S(\Delta(\rho))-S(\rho).
\end{aligned}
\end{equation}

\begin{theorem} For any  qubit pure state $|\psi\rangle$, the measure of imaginarity   based on the relative entropy  satisfies
\begin{equation}\label{6}
\begin{aligned}
M_{\rm r}(|\psi\rangle)\leq M_{l_{1}}(|\psi\rangle),
\end{aligned}
\end{equation}
the equality holds if $M_{l_{1}}(|\psi\rangle)=1$.
\end{theorem}

\textit{\textbf{Proof.}} Choose a qubit pure state $|\psi\rangle=\alpha|0\rangle+\beta|1\rangle$, where $\alpha,~\beta$ are complex numbers and satisfy $|\alpha|^{2}+|\beta|^{2}=1$.
Assume that $\alpha=c+d{\rm i},~\beta=e+f{\rm i}$, and   $H(x)=-x\log _{2}x-(1-x)\log_{2}(1-x)$. It is not difficult to obtain
\begin{equation}
\quad M_{\rm r}(|\psi\rangle)=H(\lambda_{1}),
\end{equation}
where
\begin{equation}
\lambda_{1}=\frac{1+\sqrt{1-4(cf-de)^{2}}}{2}.
\end{equation}

  According to  $H(x)\leq2\sqrt{x(1-x)}$ \cite{19}, we have
\begin{equation}\label{6}
\begin{aligned}
&\quad M_{\rm r}(|\psi\rangle)\\
&=H(\lambda_{1})\\
&\leq2\sqrt{\lambda_{1}(1-\lambda_{1})}\\
&=2\sqrt{\frac{1+\sqrt{1-4(cf-de)^{2}}}{2}\times\frac{1-\sqrt{1-4(cf-de)^{2}}}{2}}\\
&=2\sqrt{\frac{1}{4}-\frac{1}{4}(1-4(cf-de)^{2})}\\
&=2\sqrt{(cf-de)^{2}}\\
&=2|cf-de|\\
&=M_{l_{1}}(|\psi\rangle).
\end{aligned}
\end{equation}
Thus we have proved that for a qubit pure state $|\psi\rangle$,  $M_{\rm r}(|\psi\rangle)\leq M_{l_{1}}(|\psi\rangle)$ is true.

Clearly, when  $M_{l_{1}}(|\psi\rangle)=1$, one has $|cf-de|=\frac {1}{2}$. So $\lambda_1=\frac {1}{2}$, which induces $1=H(\lambda_1)=M_{\rm r}(|\psi\rangle)$. This fact shows that $M_{l_{1}}(|\psi\rangle)=M_{\rm r}(|\psi\rangle)$, if  $M_{l_{1}}(|\psi\rangle)=1$. So Theorem 4 holds.

In addition to  the above  measures of imaginarity, there exist other measures.  Next, based on the  measure of imaginarity  of pure states, we will give a measure of imaginarity of mixed quantum states by convex roof extended \cite{25}.

\begin{theorem} If $M(|\psi\rangle)$ is a  measure of imaginarity of pure state $|\psi\rangle$, then the convex roof extended
\begin{equation}\label{6}
\begin{aligned}
M(\rho)=\min_{\{p_{i},|\psi_{i}\rangle\}}\sum_{i} p_{i}M(|\psi_{i}\rangle)
\end{aligned}
\end{equation}
is  a  measure of imaginarity of mixed state  $\rho$ if $M(\rho)$ is a convex function.  Here $\{p_{i},|\psi_{i}\rangle\}$ is the decomposition of quantum state $\rho$, and $\{p_i\}$ is a probability distribution, namely, $\rho=\sum_{i}p_{i}|\psi_{i}\rangle\langle\psi_{i}|$.
\end{theorem}

\textit{\textbf{Proof.}} According to the definition of the function $M(\rho)$, when  $M(\rho)=0$, obviously we can obtain that quantum state $\rho$ is a real one. Conversely, if $\rho$ is a real state, there is a real decomposition $\rho=\sum_i p_i|\psi_i\rangle\langle\psi_i|$ such that $M(|\psi_i\rangle)=0$. So $M(\rho)=0$.

Next, we will prove that the function $M(\rho)$ is   monotonic.

For any quantum state $\rho$, we take the best decomposition of quantum state $\rho$, expressed as $\rho=\sum_{k}p_{k}|\psi_{k}\rangle\langle\psi_{k}|$, then one has
\begin{equation}\label{6}
\begin{aligned}
 M(\rho)=\sum_{k} p_{k}M(|\psi_{k}\rangle).
\end{aligned}
\end{equation}
Assume  $\{K_{j}\}$ is the set of Kraus operators of a real operation,  $c_{jk}=\langle\psi_{k}|K_{j}^{{\rm T}}K_{j}|\psi_{k}\rangle,~q_{j}={\rm Tr}K_{j}\rho K_{j}^{{\rm T}}$,  then
\begin{equation}\label{6}
\begin{aligned}
&\quad\sum_{j}q_{j}M(\frac{K_{j}\rho K_{j}^{{\rm T}}}{q_{j}})\\
&=\sum_{j}q_{j}M(\sum_{k}p_{k}\frac{K_{j}|\psi_{k}\rangle\langle\psi_{k}| K_{j}^{{\rm T}}}{q_{j}})\\
&=\sum_{j}q_{j}M(\sum_{k}\frac{p_{k}c_{jk}}{q_{j}}\times\frac{K_{j}|\psi_{k}\rangle\langle\psi_{k}| K_{j}^{{\rm T}}}{c_{jk}})\\
&\leq\sum_{j,k}p_{k}c_{jk}M(\frac{K_{j}|\psi_{k}\rangle\langle\psi_{k}| K_{j}^{{\rm T}}}{c_{jk}})\\
&\leq\sum_{k}p_{k}M(|\psi_{k}\rangle)\\
&=M(\rho),
\end{aligned}
\end{equation}
where first inequality is true because $M(\rho)$  is a convex function. Thus we demonstrate  that the function $M(\rho)$  is monotonic. So Theorem 5 holds.

\section{Influence of quantum  Channel on quantum state order}\label{sec:review}

In this section, we mainly investigate  the ordering of quantum states based on the measure of imaginarity  after passing through a real channel. The main real channels involved are amplitude damping channel, phase flip channel, and bit flip channel.  We restate the definition of the ordering of quantum states as follows \cite{24,26,27}.

\begin{definition}   Let  $M_{A}$ and $M_{B}$ be  two measures of imaginarity.  For  arbitrary  two quantum states $\rho_{1}$ and $\rho_{2}$, if the following relationship
\begin{equation}\label{6}
 M_{A}(\rho_{1})\leq M_{A}(\rho_{2})\Leftrightarrow M_{B}(\rho_{1})\leq M_{B}(\rho_{2})
\end{equation}
is true, then the  measures $M_{A}$ and $M_{B}$  are said to be of the same order, if the above relation is not satisfied, the measures $M_{A}$ and $M_{B}$   are considered to be of different order.
\end{definition}

We only discuss the ordering of quantum states in the case of a single qubit. In  a fixed reference basis, the state of a single-qubit  can always be written as
\begin{equation}\label{538}
\begin{aligned}
 \rho &=\frac{1}{2}({\mathbb {I}}+{\mathbf {r}} \cdot  {\bf{\sigma}})\\
&=\frac{1}{2}(\mathbb{I}+t{\mathbf{n}}\cdot{\bf{\sigma}})\\
&=\begin{pmatrix}
\frac{1+tn_{z}}{2}&\frac{t(n_{x}-{\rm i}n_{y})}{2}\\
\frac{t(n_{x}+{\rm i}n_{y})}{2}&\frac{1-tn_{z}}{2}
\end{pmatrix},
\end{aligned}
\end{equation}
where $\bf{\sigma}$ is the Pauli vector, $t=\|{\mathbf{r}}\|\leq 1$, ${\mathbf{n}}=(n_{x},~n_{y},~n_{z})=\frac{1}{t}{\mathbf{r}}$ is a unitary vector.

It is easy to  obtain  that the  measures of imaginarity of quantum state $\rho$
\begin{equation}\label{548}
\begin{aligned}
M_{l_{1}}(\rho)=t|n_{y}|,
\end{aligned}
\end{equation}
\begin{equation}\label{8}
\begin{aligned}
M_{{\rm r}}(\rho)=H(\frac{1}{2}+\frac{t\sqrt{1-n_{y}^{2}}}{2})-H(\frac{1+ t}{2}).
\end{aligned}
\end{equation}

 Now let's consider the monotonicity of these functions. One can easily obtain
\begin{equation}\label{856}
\begin{aligned}
\frac{\partial M_{{\rm r}}(\rho)}{\partial |n_{y}|}=\frac{t}{2}\cdot\frac{-|n_{y}|}{\sqrt{1-n_{y}^{2}}}\log_{2}\frac{1-t\sqrt{1-n_{y}^{2}}}{1+t\sqrt{1-n_{y}^{2}}}\geq0.
\end{aligned}
\end{equation}

\begin{equation}\label{8}
\begin{aligned}
\frac{\partial M_{{\rm r}}(\rho)}{\partial t}=\frac{1}{2}\log_{2}\frac{1+t}{1-t}+\frac{\sqrt{1-n_{y}^{2}}}{2}\log_{2}\frac{1-t\sqrt{1-n_{y}^{2}}}{1+t\sqrt{1-n_{y}^{2}}}.
\end{aligned}
\end{equation}
Because the function $f(x)=x\log_{2}\frac{1-tx}{1+tx}$ $(0\leq x\leq1)$ is decreasing monotonically, so we have
\begin{equation}\label{158}
\begin{aligned}
\frac{\partial M_{{\rm r}}(\rho)}{\partial t}&=\frac{1}{2}\log_{2}\frac{1+t}{1-t}+\frac{\sqrt{n_{x}^{2}+n_{z}^{2}}}{2}\log_{2}\frac{1-t\sqrt{n_{x}^{2}+n_{z}^{2}}}{1+t\sqrt{n_{x}^{2}+n_{z}^{2}}}\\
&\geq\frac{1}{2}\log_{2}\frac{1+t}{1-t}+\frac{1}{2}\log_{2}\frac{1-t}{1+t}\\
&\geq0.
\end{aligned}
\end{equation}
Therefore, $M_{{\rm r}}(\rho)$ is monotonic increasing about the independent variables  $|n_{y}|$ and $t$. Evidently $M_{l_{1}}(\rho)$  is also monotonic increasing about the independent variables  $|n_{y}|$ and $t$.  Thus we have the following conclusion.

\begin{proposition}
 The measure $M_{l_{1}}(\rho)$ and the measure $M_{{\rm r}}(\rho)$ are of the same order for qubit quantum states.
\end{proposition}

It is well known that the quantum channel can change the quantum state, furthermore it can affect the quantum state order also. For a  measure  of quantum states, we define the influence of quantum channel on quantum state order as follows.

\begin{definition}   Let  $M$  be a measure of imaginarity and $\varepsilon$ be a quantum channel.  For  arbitrary two  quantum states $\rho_{1}$ and $\rho_{2}$, if
\begin{equation}\label{6}
 M(\rho_{1})\leq M(\rho_{2})\Leftrightarrow M(\varepsilon(\rho_{1}))\leq M(\varepsilon(\rho_{2}))
\end{equation}
holds, then we say the quantum channel $\varepsilon$ does not change the quantum state order; otherwise we say the quantum state order is changed by the quantum channel $\varepsilon$.
\end{definition}

Next, we discuss the influence of a quantum  channels on the ordering of qubit quantum states when one chooses a  measure of imaginarity. Firstly we study the case of  the  bit flip channel $\varepsilon$ and imaginarity measure $M_{\rm r}(\rho)$. Here the quantum state of the qubit is  stated as Eq.(\ref{538}),          the  bit flip channel $\varepsilon$ is expressed by the real Kraus operators  $\{K_{0}=\sqrt{p}\mathbb{I}, ~K_{1}=\sqrt{1-p}\sigma_{x}\}$, where $p\in[0,1]$, $\sigma_x$ is the Pauli operator.

\begin{proposition} Suppose one chooses $M_{\rm r}(\rho)$ as the measure of imaginarity, then the quantum state order does not change  after a single-qubit  goes through a bit flip channel.
\end{proposition}

\textit{\textbf{Proof.}} The  state of the qubit system after passing through the bit flip channel $\varepsilon$ is
\begin{equation}\label{860}
\begin{aligned}
\varepsilon(\rho)&=K_{0}\rho K_{0}^{\dagger}+K_{1}\rho K_{1}^{\dagger}\\
&=\begin{pmatrix}
\frac{1+tn_{z}(2p-1)}{2}&\frac{tn_{x}-{\rm i}tn_{y}(2p-1)}{2}\\
\frac{tn_{x}+{\rm i}tn_{y}(2p-1)}{2}&\frac{1-tn_{z}(2p-1)}{2}
\end{pmatrix},
\end{aligned}
\end{equation}
where $\rho$ is  expressed by Eq.(\ref{538}).

It is easy to derive that
\begin{equation}\label{8}
\begin{aligned}
M_{\rm r}(\varepsilon(\rho))&=H(\frac{1+t\sqrt{n_{x}^{2}+(2p-1)^{2}n_{z}^{2}}}{2})\\
&\quad -H(\frac{1+t\sqrt{n_{x}^{2}+(2p-1)^{2}(1-n_{x}^{2})}}{2}).
\end{aligned}
\end{equation}
Obviously,  $M_{\rm r}(\varepsilon(\rho))$ contains  four parameters $t,p,n_{x},n_{z}$. We can easily get
\begin{equation}\label{628}
\begin{aligned}
&\quad\frac{\partial M_{\rm r}(\varepsilon(\rho))}{\partial |n_{z}|}\\
&=\frac{t}{2}\cdot\frac{(2p-1)^{2}|n_{z}|}{\sqrt{n_{x}^{2}+(2p-1)^{2}n_{z}^{2}}}\log_{2}\frac{1-t\sqrt{n_{x}^{2}+(2p-1)^{2}n_{z}^{2}}}{1+t\sqrt{n_{x}^{2}+(2p-1)^{2}n_{z}^{2}}}\\
&\leq0.
\end{aligned}
\end{equation}
By  using the monotonically increasing property of
\begin{equation}\label{8}
\begin{aligned}
f(x)=\frac{1}{x}\log_{2}\frac{1+tx}{1-tx},(0\leq x\leq1),
\end{aligned}
\end{equation}
 we have
\begin{equation}\label{8}
\begin{aligned}
&\quad \frac{\partial M_{\rm r}(\varepsilon(\rho))}{\partial |n_{x}|}\\
&=\frac{t}{2}\cdot\frac{|n_{x}|}{\sqrt{n_{x}^{2}+(2p-1)^{2}n_{z}^{2}}}\cdot\log_{2}\frac{1-t\sqrt{n_{x}^{2}+(2p-1)^{2}n_{z}^{2}}}{1+t\sqrt{n_{x}^{2}+(2p-1)^{2}n_{z}^{2}}}\\
&\quad-\frac{t}{2}\cdot\frac{|n_{x}|[1-(2p-1)^{2}]}{\sqrt{n_{x}^{2}+(2p-1)^{2}n_{y}^{2}+(2p-1)^{2}n_{z}^{2}}}\cdot\\
&\quad\log_{2}\frac{1-t\sqrt{n_{x}^{2}+(2p-1)^{2}n_{z}^{2}+(2p-1)^{2}n_{y}^{2}}}{1+t\sqrt{n_{x}^{2}+(2p-1)^{2}n_{z}^{2}+(2p-1)^{2}n_{y}^{2}}}\\
&\geq\frac{t}{2}\cdot\frac{|n_{x}|(2p-1)^{2}}{\sqrt{n_{x}^{2}+(2p-1)^{2}n_{z}^{2}}}\log_{2}\frac{1-t\sqrt{n_{x}^{2}+(2p-1)^{2}n_{z}^{2}}}{1+t\sqrt{n_{x}^{2}+(2p-1)^{2}n_{z}^{2}}}.
\end{aligned}
\end{equation}
So when  $n_{x}\leq 0$, we have  $\frac{\partial M_{\rm r}(\varepsilon(\rho))}{\partial n_{x}}\geq0$. Because $M_{\rm r}(\varepsilon(\rho))$ is an even function of the variable $n_{x}$, we can  conclude that  $ M_{\rm r}(\varepsilon(\rho))$ is  monotonic decreasing function of variable $|n_{x}|$, i.e.
\begin{equation}\label{658}
\frac{\partial M_{\rm r}(\varepsilon(\rho))}{\partial |n_{x}|}\leq 0.
\end{equation}

The partial derivative of $M_{{\rm r}}(\varepsilon(\rho))$ with respect to $t$ is
\begin{equation}\label{6668}
\begin{aligned}
&\quad \frac{\partial M_{{\rm r}}(\varepsilon(\rho))}{\partial t}\\
&=\frac{\sqrt{n_{x}^{2}+(2p-1)^{2}n_{z}^{2}}}{2}\log_{2}\frac{1-t\sqrt{n_{x}^{2}+(2p-1)^{2}n_{z}^{2}}}{1+t\sqrt{n_{x}^{2}+(2p-1)^{2}n_{z}^{2}}}\\
&\quad+\frac{\sqrt{n_{x}^{2}+(2p-1)^{2}n_{z}^{2}+(2p-1)^{2}n_{z}^{2}}}{2}\cdot\\
&\quad \log_{2}\frac{1+t\sqrt{n_{x}^{2}+(2p-1)^{2}n_{z}^{2}+(2p-1)^{2}n_{y}^{2}}}{1-t\sqrt{n_{x}^{2}+(2p-1)^{2}n_{z}^{2}+(2p-1)^{2}n_{y}^{2}}}\\
&\geq0.
\end{aligned}
\end{equation}

So the measure $M_{{\rm r}}(\varepsilon(\rho))$ is a monotonically decreasing function with respect to the variable $|n_{x}|,|n_{z}|$, and a monotonically increasing function with respect to the variable $t$.

On the other hand, we can obtain that
\begin{equation}\label{8}
\begin{aligned}
&\quad \frac{\partial M_{\rm r}(\rho)}{\partial |n_{x}|}\\
&= \frac{\partial M_{\rm r}(\rho)}{\partial |n_{y}|}\frac {\partial |n_{y}|}{\partial |n_{x}|}\\
&= \frac{\partial M_{\rm r}(\rho)}{\partial |n_{y}|}\frac {\partial \sqrt{1-n_{x}^2-n_{z}^2}}{\partial |n_{x}|}\\
&= \frac{\partial M_{\rm r}(\rho)}{\partial |n_{y}|}\frac { -|n_{x}| }{\sqrt{1-n_{x}^2-n_{z}^2}}.\\
\end{aligned}
\end{equation}
By using Eq.(\ref{856}), one gets
\begin{equation}\label{6868}
\frac{\partial M_{\rm r}(\rho)}{\partial |n_{x}|}\leq 0.
\end{equation}
Similarly, we have
\begin{equation}\label{6969}
\frac{\partial M_{\rm r}(\rho)}{\partial |n_{z}|}\leq 0.
\end{equation}
Combining Eqs. (\ref{158}), (\ref{628}), (\ref{658}), (\ref{6668}), (\ref{6868}), and (\ref{6969}), one arrives at that the quantum state order does not change  after a single-qubit   goes through a bit flip channel. Thus Proposition 2 is true.

\begin{proposition}  Assume we choose $M_{l_{1}}(\rho)$ as the measure of imaginarity, then the quantum state order does not change  after a single-qubit  goes through a bit flip channel.
\end{proposition}

\textit{\textbf{Proof.}} By using Eq.(\ref{860}) we  have
\begin{equation}\label{8}
\begin{aligned}
M_{l_{1}}(\varepsilon(\rho))=t|(2p-1)n_{y}|.
\end{aligned}
\end{equation}
Considering the above equation and Eq.(\ref{548}), it is not difficult to obtain  that when we choose $M_{l_{1}}(\rho)$ as the measure of imaginarity,  the quantum state order does not change  after  a single-qubit goes through a bit flip channel. This implies that Proposition 3 holds.

  Now let us investigate  the case that when the imaginarity measure $M_{\rm r}(\rho)$  has been choosed, and the quantum channel is the phase flip channel $\Lambda$. Here the quantum state of the qubit is  stated as Eq.(\ref{538}),          the  phase flip channel $\Lambda$ is expressed by the real Kraus operators  $K_{0}=\sqrt{p}\mathbb{I},K_{1}=\sqrt{1-p}|0\rangle\langle 0|,K_{2}=\sqrt{1-p}|1\rangle\langle 1|, 0\leq p\leq 1$.

For this case we will prove  the following proposition.

\begin{proposition} Suppose we choose $M_{\rm r}(\rho)$ as the measure of imaginarity, then the quantum state order does not change  after a single-qubit  goes through a phase flip channel.
\end{proposition}

\textit{\textbf{Proof.}} After a qubit  passes through a phase flip channel, the quantum state can be written  as
\begin{equation}\label{72728}
\begin{aligned}
\Lambda(\rho)&=K_{0}\rho K_{0}^{\dagger}+K_{1}\rho K_{1}^{\dagger}+K_{2}\rho K_{2}^{\dagger}\\
&=\begin{pmatrix}\frac{1+tn_{z}}{2}&\frac{tp(n_{x}-{\rm i}n_{y})}{2}\\
\frac{tp(n_{x}+{\rm i}n_{y})}{2}&\frac{1-tn_{z}}{2}
\end{pmatrix}.
\end{aligned}
\end{equation}
One can easily deduce
\begin{equation}\label{8}
\begin{aligned}
M_{{\rm r}}(\Lambda(\rho))&=H(\frac{1+t\sqrt{n_{z}^{2}+p^{2}n_{x}^{2}}}{2})\\
&\quad-H(\frac{1+t\sqrt{n_{z}^{2}+p^{2}(1-n_{z}^{2})}}{2}).
\end{aligned}
\end{equation}

The partial derivatives are
\begin{equation}\label{748}
\begin{aligned}
&\quad\frac{\partial M_{\rm r}(\Lambda(\rho))}{\partial t}\\
&=\frac{\sqrt{p^{2}n_{x}^{2}+n_{z}^{2}}}{2}\log_{2}\frac{1-t\sqrt{n_{z}^{2}+p^{2}n_{x}^{2}}}{1+t\sqrt{n_{z}^{2}+p^{2}n_{x}^{2}}}\\
&\quad+\frac{\sqrt{n_{z}^{2}+p^{2}(1-n_{z}^{2})}}{2}\log_{2}\frac{1+t\sqrt{n_{z}^{2}+p^{2}(1-n_{z}^{2})}}{1-t\sqrt{n_{z}^{2}+p^{2}(1-n_{z}^{2})}}\\
&\geq 0;
\end{aligned}
\end{equation}

\begin{equation}\label{758}
\begin{aligned}
&\quad\frac{\partial M_{{\rm r}}(\Lambda(\rho))}{\partial |n_{x}|}\\
&=\frac{t}{2}\cdot\frac{p^{2}|n_{x}|}{\sqrt{n_{z}^{2}+p^{2}n_{x}^{2}}}\log_{2}\frac{1-t\sqrt{n_{z}^{2}+p^{2}n_{x}^{2}}}{1+t\sqrt{n_{z}^{2}+p^{2}n_{x}^{2}}}\\
&\leq 0;
\end{aligned}
\end{equation}

\begin{equation}\label{88888}
\begin{aligned}
&\quad\frac{\partial M_{{\rm r}}(\Lambda(\rho))}{\partial |n_{z}|}\\
&=\frac{t}{2}\cdot\frac{|n_{z}|}{\sqrt{n_{z}^{2}+p^{2}n_{x}^{2}}}\log_{2}\frac{1-t\sqrt{n_{z}^{2}+p^{2}n_{x}^{2}}}{1+t\sqrt{n_{z}^{2}+p^{2}n_{x}^{2}}}\\
&\quad-\frac{t}{2}\cdot\frac{|n_{z}|(1-p^{2})} {\sqrt{n_{z}^{2}+p^{2}(1-n_{z}^{2})}}\log_{2}\frac{1-t\sqrt{n_{z}^{2}+p^{2}(1-n_{z}^{2})}}{1+t\sqrt{n_{z}^{2}+p^{2}(1-n_{z}^{2})}}\\
&\leq \frac{t}{2}\cdot\frac{|n_{z}|}{\sqrt{n_{z}^{2}}}\log_{2}\frac{1-t\sqrt{n_{z}^{2}}}{1+t\sqrt{n_{z}^{2}}}\\
&\quad-\frac{t}{2}\cdot\frac{|n_{z}|(1-p^{2})} {\sqrt{n_{z}^{2}+p^{2}(1-n_{z}^{2})}}\log_{2}\frac{1-t\sqrt{n_{z}^{2}+p^{2}(1-n_{z}^{2})}}{1+t\sqrt{n_{z}^{2}+p^{2}(1-n_{z}^{2})}}\\
&\leq \frac{t}{2}\cdot\frac{|n_{z}|}{\sqrt{n_{z}^{2}}}\log_{2}\frac{1-t\sqrt{n_{z}^{2}}}{1+t\sqrt{n_{z}^{2}}}-\frac{t}{2}\cdot\frac{|n_{z}|}{\sqrt{n_{z}^{2}}}\log_{2}\frac{1-t\sqrt{n_{z}^{2}}}{1+t\sqrt{n_{z}^{2}}}\\
&= 0.
\end{aligned}
\end{equation}
Here the first inequality in Eq.(\ref{88888}) comes from the fact that when $t$ is fixed,  $ \frac {1}{x}\log_2\frac {1-tx}{1+tx}$ is  monotonically decreasing function with respect to $x$  and $n_{z}^{2}+n_{x}^{2}\leq 1$;  the second  inequality in Eq.(\ref{88888}) is based on that when $n_z, t$ are fixed,  $-\frac{t}{2}\cdot\frac{|n_{z}|(1-p^{2})} {\sqrt{n_{z}^{2}+p^{2}(1-n_{z}^{2})}}\log_{2}\frac{1-t\sqrt{n_{z}^{2}+p^{2}(1-n_{z}^{2})}}{1+t\sqrt{n_{z}^{2}+p^{2}(1-n_{z}^{2})}}\\$ is monotonically decreasing function with respect to $p^2$.

By Eqs. (\ref{158}), (\ref{6868}), (\ref{6969}), (\ref{748}), (\ref{758}), (\ref{88888}) we obtain that if we choose $M_{\rm r}(\rho)$ as the measure of imaginarity, then the quantum state order does not change  after a single-qubit  goes through a phase flip channel. Thus Proposition 4 is true.

\begin{proposition}
Suppose we choose $M_{l_1}(\rho)$ as the measure of imaginarity, then the quantum state order does not change  after a single-qubit  goes through a phase flip channel.
\end{proposition}

\textit{\textbf{Proof.}} By using Eq.(\ref{72728}), we have
\begin{equation}\label{8}
\begin{aligned}
M_{l_{1}}(\Lambda(\rho))=tp|n_{y}|.
\end{aligned}
\end{equation}
By considering Eq.(\ref{548}) and above equation one can easily see that Proposition 5 holds.

 Next we discuss the influence of an amplitude damping channel on the ordering of quantum states. Here an amplitude damping channel $\Gamma$ is expressed by the real Kraus operators  $\{K_{0}=|0\rangle\langle0|+\sqrt{1-p}|1\rangle\langle1|,~K_{1}=\sqrt{p}|0\rangle\langle 1|, ~0\leq p\leq 1\}$. We will prove the following result.

\begin{proposition} When   qubit state $\rho$  satisfies $n_{z}\leq0$,  if one  chooses $M_{\rm r}(\rho)$ as the measure of imaginarity, then the quantum state order does not change  after a single-qubit  goes through an amplitude damping channel
\end{proposition}

\textit{\textbf{Proof.}} For a qubit state stated by Eq.(\ref{538}), the amplitude damping channel leads it to
\begin{equation}\label{788}
\begin{aligned}
\Gamma(\rho)&=K_{0}\rho K_{0}^{\dagger}+K_{1}\rho K_{1}^{\dagger}\\
&=\begin{pmatrix}\frac{1+tn_{z}}{2}+\frac{p(1-tn_{z})}{2}&\frac{\sqrt{1-p}t(n_{x}-{\rm i}n_{y})}{2}\\
\frac{\sqrt{1-p}t(n_{x}+{\rm i}n_{y})}{2}&\frac{(1-p)(1-tn_{z})}{2}
\end{pmatrix}.
\end{aligned}
\end{equation}

One can easily obtain the measure of imaginarity   based on relative entropy
\begin{equation}\label{8}
\begin{aligned}
&\quad M_{\rm r}(\Gamma(\rho))\\
&=H(\frac{1+\sqrt{[p+tn_{z}(1-p)]^{2}+(1-p)t^{2}n_{x}^{2}}}{2})\\
&\quad-H(\frac{1+\sqrt{[p+tn_{z}(1-p)]^{2}+(1-p)t^{2}(1-n_{z}^{2})}}{2}).
\end{aligned}
\end{equation}

Therefore, we get  the partial derivatives
\begin{equation}\label{808}
\begin{aligned}
&\quad \frac{\partial M_{\rm r}(\Gamma(\rho))}{\partial t}\\
&=\frac{[p+tn_{z}(1-p)]n_{z}(1-p)+t(1-p)n_{x}^{2}}{2\sqrt{[p+tn_{z}(1-p)]^{2}+(1-p)t^{2}n_{x}^{2}}}\\
&\quad\times\log_{2}\frac{1-\sqrt{[p+tn_{z}(1-p)]^{2}+(1-p)t^{2}n_{x}^{2}}}{1+\sqrt{[p+tn_{z}(1-p)]^{2}+(1-p)t^{2}n_{x}^{2}}}\\
&\quad+\frac{[p+tn_{z}(1-p)]n_{z}(1-p)+t(1-p)(n_{x}^{2}+n_{y}^{2})}{2\sqrt{[p+tn_{z}(1-p)]^{2}+(1-p)t^{2}(n_{x}^{2}+n_{y}^{2})}}\\
&\quad\times\log_{2}\frac{1+\sqrt{[p+tn_{z}(1-p)]^{2}+(1-p)t^{2}(n_{x}^{2}+n_{y}^{2})}}{1-\sqrt{[p+tn_{z}(1-p)]^{2}+(1-p)t^{2}(n_{x}^{2}+n_{y}^{2})}}\\
&\geq 0;
\end{aligned}
\end{equation}
\begin{equation}\label{818}
\begin{aligned}
&\quad\frac{\partial M_{\rm r}(\Gamma(\rho))}{\partial |n_{x}|}\\
&=\frac{(1-p)t^{2}|n_{x}|}{2\sqrt{[p+tn_{z}(1-p)]^{2}+(1-p)t^{2}n_{x}^{2}}}\\
&\quad \times\log_{2}\frac{1-\sqrt{[p+tn_{z}(1-p)]^{2}+(1-p)t^{2}n_{x}^{2}}}{1+\sqrt{[p+tn_{z}(1-p)]^{2}+(1-p)t^{2}n_{x}^{2}}}\\
&\leq0;
\end{aligned}
\end{equation}

\begin{equation}\label{828}
\begin{aligned}
&\quad\frac{\partial M_{\rm r}(\Gamma(\rho))}{\partial n_{z}}\\
&=\frac{[p+tn_{z}(1-p)]t(1-p)}{2\sqrt{[p+tn_{z}(1-p)]^{2}+(1-p)t^{2}n_{x}^{2}}}\\
&\quad\times\log_{2}\frac{1-\sqrt{[p+tn_{z}(1-p)]^{2}+(1-p)t^{2}n_{x}^{2}}}{1+\sqrt{[p+tn_{z}(1-p)]^{2}+(1-p)t^{2}n_{x}^{2}}}\\
&\quad+\frac{[p+tn_{z}(1-p)]t(1-p)-(1-p)t^{2}n_{z}}{2\sqrt{[p+tn_{z}(1-p)]^{2}+(1-p)t^{2}(1-n_{z}^{2})}}\\
&\quad\times\log_{2}\frac{1+\sqrt{[p+tn_{z}(1-p)]^{2}+(1-p)t^{2}(1-n_{z}^{2})}}{1-\sqrt{[p+tn_{z}(1-p)]^{2}+(1-p)t^{2}(1-n_{z}^{2})}}.
\end{aligned}
\end{equation}

By  using the monotonically increasing property of
\begin{equation}\label{8}
\begin{aligned}
f(x)=\frac{1}{x}\log_{2}\frac{1+x}{1-x},(0\leq x\leq1),
\end{aligned}
\end{equation}
and $0\leq n_x^2\leq 1-n_z^2$, then we have
\begin{equation}\label{828}
\begin{aligned}
&\quad\frac{\partial M_{\rm r}(\Gamma(\rho))}{\partial n_{z}}\\
&\geq {\rm Min}\bigg\{\frac{[p+tn_{z}(1-p)]t(1-p)}{2\sqrt{[p+tn_{z}(1-p)]^{2}}}\\
&\quad\times\log_{2}\frac{1-\sqrt{[p+tn_{z}(1-p)]^{2}}}{1+\sqrt{[p+tn_{z}(1-p)]^{2}}}\\
&\quad+\frac{[p+tn_{z}(1-p)]t(1-p)-(1-p)t^{2}n_{z}}{2\sqrt{[p+tn_{z}(1-p)]^{2}+(1-p)t^{2}(1-n_{z}^{2})}}\\
&\quad\times\log_{2}\frac{1+\sqrt{[p+tn_{z}(1-p)]^{2}+(1-p)t^{2}(1-n_{z}^{2})}}{1-\sqrt{[p+tn_{z}(1-p)]^{2}+(1-p)t^{2}(1-n_{z}^{2})}}, \\
&\quad\frac{(1-p)t^{2}n_{z}}{2\sqrt{[p+tn_{z}(1-p)]^{2}+(1-p)t^{2}(1-n_{z}^{2})}}\\
&\quad\times\log_{2}\frac{1-\sqrt{[p+tn_{z}(1-p)]^{2}+(1-p)t^{2}(1-n_{z}^{2})}}{1+\sqrt{[p+tn_{z}(1-p)]^{2}+(1-p)t^{2}(1-n_{z}^{2})}}\bigg\}.
\end{aligned}
\end{equation}
Let
\begin{equation}
\begin{aligned}
A=&\frac{[p+tn_{z}(1-p)]t(1-p)}{2\sqrt{[p+tn_{z}(1-p)]^{2}}}\\
&\quad\times\log_{2}\frac{1-\sqrt{[p+tn_{z}(1-p)]^{2}}}{1+\sqrt{[p+tn_{z}(1-p)]^{2}}}\\
&\quad+\frac{[p+tn_{z}(1-p)]t(1-p)-(1-p)t^{2}n_{z}}{2\sqrt{[p+tn_{z}(1-p)]^{2}+(1-p)t^{2}(1-n_{z}^{2})}}\\
&\quad\times\log_{2}\frac{1+\sqrt{[p+tn_{z}(1-p)]^{2}+(1-p)t^{2}(1-n_{z}^{2})}}{1-\sqrt{[p+tn_{z}(1-p)]^{2}+(1-p)t^{2}(1-n_{z}^{2})}}, \\
\end{aligned}\end{equation}
\begin{equation}
\begin{aligned}
B=&\frac{(1-p)t^{2}n_{z}}{2\sqrt{[p+tn_{z}(1-p)]^{2}+(1-p)t^{2}(1-n_{z}^{2})}}\\
&\quad\times\log_{2}\frac{1-\sqrt{[p+tn_{z}(1-p)]^{2}+(1-p)t^{2}(1-n_{z}^{2})}}{1+\sqrt{[p+tn_{z}(1-p)]^{2}+(1-p)t^{2}(1-n_{z}^{2})}}.
\end{aligned}\end{equation}
\begin{equation}
\begin{aligned}
&\quad A-B\\
&=\frac{[p+tn_{z}(1-p)]t(1-p)}{2\sqrt{[p+tn_{z}(1-p)]^{2}}}\\
&\quad\times\log_{2}\frac{1-\sqrt{[p+tn_{z}(1-p)]^{2}}}{1+\sqrt{[p+tn_{z}(1-p)]^{2}}}\\
&\quad+\frac{[p+tn_{z}(1-p)]t(1-p)}{2\sqrt{[p+tn_{z}(1-p)]^{2}+(1-p)t^{2}(1-n_{z}^{2})}}\\
&\quad\times\log_{2}\frac{1+\sqrt{[p+tn_{z}(1-p)]^{2}+(1-p)t^{2}(1-n_{z}^{2})}}{1-\sqrt{[p+tn_{z}(1-p)]^{2}+(1-p)t^{2}(1-n_{z}^{2})}}. \\
\end{aligned}
\end{equation}
So when $p+tn_{z}(1-p)\geq 0$, we have
$A\geq B$;  when $p+tn_{z}(1-p)\leq 0$, we have
$A\leq B$.

In the situation $p+tn_{z}(1-p)\leq 0$, one gets
\begin{equation}\label{898}
\begin{aligned}
&\quad\frac{\partial M_{\rm r}(\Gamma(\rho))}{\partial n_{z}}\\
&\geq A\\
&=\frac{-t(1-p)}{2}\log_{2}\frac{1-\sqrt{[p+tn_{z}(1-p)]^{2}}}{1+\sqrt{[p+tn_{z}(1-p)]^{2}}}\\
&\quad+\frac{t(1-p)}{2}\frac{p(1-tn_{z})}{\sqrt{[p+tn_{z}(1-p)]^{2}+(1-p)t^{2}(1-n_{z}^{2})}}\\
&\quad\times\log_{2}\frac{1+\sqrt{[p+tn_{z}(1-p)]^{2}+(1-p)t^{2}(1-n_{z}^{2})}}{1-\sqrt{[p+tn_{z}(1-p)]^{2}+(1-p)t^{2}(1-n_{z}^{2})}}\\
&\geq 0.\\
\end{aligned}
\end{equation}

On the other hand, in the case $p+tn_{z}(1-p)\geq 0$ we have
\begin{equation}\label{888}
\begin{aligned}
&\quad\frac{\partial M_{\rm r}(\Gamma(\rho))}{\partial n_{z}}\\
&\geq B\\
&=\frac{(1-p)t^{2}n_{z}}{2\sqrt{[p+tn_{z}(1-p)]^{2}+(1-p)t^{2}(1-n_{z}^{2})}}\\
&\quad\times\log_{2}\frac{1-\sqrt{[p+tn_{z}(1-p)]^{2}+(1-p)t^{2}(1-n_{z}^{2})}}{1+\sqrt{[p+tn_{z}(1-p)]^{2}+(1-p)t^{2}(1-n_{z}^{2})}}.
\end{aligned}
\end{equation}
So when $n_{z}\leq0$ and  $p+tn_{z}(1-p)\geq 0$, we have $\frac{\partial M_{\rm r}(\Gamma(\rho))}{\partial n_{z}}\geq0$, that is, if $n_x, t, p$ are fixed and satisfy $n_{z}\leq0$ and  $p+tn_{z}(1-p)\geq 0$, then  the function $ M_{\rm r}(\Gamma(\rho))$ is monotonically increasing with respect to the variables $n_{z}$.

Combining Eqs. (\ref{158}),  (\ref{6868}), (\ref{6969}), (\ref{808}), (\ref{818}), (\ref{898}), (\ref{888}), we arrive at that  when  qubit state $\rho$  satisfies $n_{z}\leq0$,  if one  chooses $M_{\rm r}(\rho)$ as the measure of imaginarity, then the quantum state order does not change  after a single-qubit  goes through an amplitude damping channel. Thus we have demonstrated Proposition 6.

\begin{proposition} When we  take $M_{l_1}(\rho)$ as the measure of imaginarity, then the quantum state order does not change  after a single-qubit  goes through an amplitude damping channel.
\end{proposition}

\textit{\textbf{Proof.}} By using Eq.(\ref{788}) we can easily deduce the  measure of imaginarity
\begin{equation}\label{8}
\begin{aligned}
M_{l_{1}}(\Gamma(\rho))=t\sqrt{1-p}|n_{y}|.
\end{aligned}
\end{equation}

By using Eq.(\ref{548}) and above equation one can easily obtain  that Proposition 7 is true.

\section{conclusion}\label{sec:conclusion}

In summary, we study the measures of imaginarity in the framework of resource theory and the quantum state order after a quantum system passes through a real channel. We define functions based on $l_{1}$ norm and the convex roof extended, and  show that they are the measures of imaginarity.
The  relationships between relative entropy of imaginarity  $M_{\rm r}(\rho)$  and  the imaginarity measure $M_{l_{1}}(\rho)$ based on $l_{1}$ norm for the single-qubit pure state $\rho$ is investigated. We also prove that the functions based on $l_{p}$ norm and $p-$norm are not the measures of  imaginarity. Moreover, we demonstrate that  the measure $M_{l_{1}}(\rho)$ and the measure $M_{{\rm r}}(\rho)$ are of the same order for qubit quantum states and  discuss the influences of the bit flip channel, phase damping channel and amplitude flip channel on single-qubit  state order, respectively.

\begin{acknowledgments}
This work was supported by the National Natural Science Foundation of China under Grant Nos. 62271189, 12071110,  the Hebei Natural Science Foundation of China under Grant No. A2020205014,  funded by Science and Technology Project of Hebei Education Department under Grant Nos. ZD2020167, ZD2021066, and  the Hebei Central Guidance on Local Science and Technology Development Foundation of China under Grant No. 226Z0901G.

\end{acknowledgments}


\end{document}